\documentclass{article}
\usepackage[T1]{fontenc}
\usepackage[latin9]{inputenc}
\usepackage{color}
\usepackage{graphicx}
\usepackage{esint}

\makeatletter

\providecommand{\tabularnewline}{\\}

\makeatother

\begin{document}

\title{\textcolor{black}{Testing quantised inertia on the emdrive}}

\author{\textcolor{black}{M.E. McCulloch}%
\thanks{\textcolor{black}{SMSE, Plymouth University, Plymouth, PL4 8AA. mike.mcculloch@plymouth.ac.uk}%
}}
\maketitle
\begin{abstract}
\textcolor{black}{It has been shown that truncated cone-shaped cavities
with microwaves resonating within them move slightly towards their
narrow ends (the emdrive). Standard physics has no explanation for
this and an error has not yet been found. It is shown here that this
effect can be predicted by assuming that the inertial mass of the
photons in the cavity is caused by Unruh radiation, whose wavelengths
must fit exactly within the cavity, using a theory already applied
successfully to astrophysical anomalies such as galaxy rotation where
the Unruh waves have to fit within the Hubble scale. In the emdrive
this means that more Unruh waves are allowed at the wide end, leading
to a greater inertial mass for the photons there, and to conserve
momentum the cavity must move towards its narrow end, as observed.
The model predicts thrusts of: 3.8, 149, 7.3, 0.23, 0.57, 0.11, 0.64
and 0.02 mN compared with the observed thrusts of: 16, 147, 9, 0.09,
0.05, 0.06, 0.03, and 0.02 mN and predicts that if the axial length
is equal to the diameter of the small end of the cavity, the thrust
should be reversed.}
\end{abstract}

\section{\textcolor{black}{Introduction}}

\textcolor{black}{It was first demonstrated by Shawyer (2008) that
when microwaves resonate within a truncated cone-shaped cavity a small,
unexplained thrust and acceleration occurs towards the narrow end.
In one example when 850W of power was put into such a cavity with
end diameters of 16 cm and 12cm and a Q value (dissipation constant)
of 5900 the thrust measured was 16mN towards the narrow end. The results
from two of Shawyer's experiments are shown in Table 1 (rows 1-2).
There is no explanation for this behaviour in standard physics because
it violates the conservation of momentum, and Shawyer's own attempt
to explain it using special relativity is not convincing, as this
theory also should obey the conservation of momentum (Mullins, 2006).}

\textcolor{black}{Nethertheless, this anomaly has been confirmed by
a Chinese team (Juan et al., 2012) who put 80-2500W of power into
a similar cavity at a frequency of 2.45GHz and measured a thrust of
between 70mN and 720mN. Their result cannot however be used for testing
here since they did not specify their cavity's quality factor or their
geometry. A similar anomaly was seen by Fetta (2012) who claimed the
effect was due to grooves in one of the ends of the cavity and not
the asymmetry of the cone. A later test by NASA showed the grooves
made no difference to the thrust.}

\textcolor{black}{Further confirmation of the emdrive was obtained
by a NASA team (Brady et al., 2014), most recently in a vacuum proving
that the effect is not due to moving air. Their four results are shown
in Table 1 (rows 4 to 7). They did provide details of their Q factor
and some details of their cavity's geometry.}

\textcolor{black}{McCulloch (2007) has proposed a new model for inertia
(MiHsC) that assumes that the inertia of an object is due to the Unruh
radiation it sees when it accelerates, radiation which is also subject
to a Hubble-scale Casimir effect. In this model only Unruh wavelengths
that fit exactly into twice the Hubble diameter are allowed, so that
a greater proportion of the waves are disallowed for low accelerations
(which see longer Unruh waves) leading to a gradual new loss of inertia
as accelerations become tiny. MiHsC modifies the standard inertial
mass ($m$) to a modified one ($m_{i}$) as follows:}

\textcolor{black}{
\begin{equation}
m_{i}=m\left(1-\frac{2c^{2}}{|a|\Theta}\right)=m\left(1-\frac{\lambda}{4\Theta}\right)
\end{equation}
}

\textcolor{black}{where c is the speed of light, $\Theta$ is twice
the Hubble distance, '|a|' is the magnitude of the relative acceleration
of the object relative to surrounding matter and $\lambda$ is the
peak wavelength of the Unruh radiation it sees. Eq. 1 predicts that
for terrestrial accelerations (eg: $9.8m/s^{2}$) the second term
in the bracket is tiny and standard inertia is recovered, but in low
acceleration environments, for example at the edges of galaxies (when
a is small and $\lambda$ is large) the second term in the bracket
becomes larger and the inertial mass decreases in a new way so that
MiHsC can explain galaxy rotation without the need for dark matter
(McCulloch, 2012) and cosmic acceleration without the need for dark
energy (McCulloch, 2007, 2010). However, astrophysical tests like
these can be ambiguous, since more flexible theories like dark matter
can be manipulated to fit the data, and so a controlled laboratory
test like the EmDrive is preferable.}

\textcolor{black}{The difficulty of demonstrating MiHsC on Earth is
the huge size of $\Theta$ in Eq. 1 which makes the effect very small
unless the acceleration is tiny, as in deep space. One way to make
the effect more obvious is to reduce the distance to the horizon $\Theta$
and this is what the emdrive may be doing since the radiation within
it is accelerating so fast that the Unruh waves it sees will be short
enough to be limited by the cavity walls in a MiHsC-like manner. McCulloch
(2015) showed that assuming that the inertial mass of the photons
is determined by MiHsC and the width of the cavity, and assuming the
conservation of momentum, predicts a new force of size}

\textcolor{black}{
\begin{equation}
F=-\frac{PQL}{c}\left(\frac{1}{w_{s}}-\frac{1}{w_{b}}\right)
\end{equation}
}

\textcolor{black}{where P is the power input, Q is the quality factor,
L is the axial length, c is the speed of light and $w_{s}$ and $w_{b}$
are the widths of the small and big ends respectively. This formula
predicts the observed emdrive thrusts quite well (see the Table, column
6) however it assumed that the Unruh waves only resonate perpendicular
to the axis of symmetry and so predicts infinite thrust for a pointed
cone ($w_{s}=0)$. In this paper this model is modified to approximate
the microwaves' resonance in three-dimensions.}

\section{\textcolor{black}{Method}}

\textcolor{black}{The so-called emdrive is a microwave resonant cavity
shaped like a truncated cone, with one round end larger than the other
(see Figure 1). When an electromagnetic field resonates in the cavity
we can consider the conservation of momentum for the photons as they
move along the axis}

\textcolor{black}{
\begin{equation}
\frac{\partial(mv)}{\partial t}=0=m\frac{\partial v}{\partial t}+v\frac{\partial m}{\partial t}
\end{equation}
}

\textcolor{black}{The first term on the right hand side is the force
(mass times acceleration) that must be exerted on the photons to change
their speed to conserve their momentum if MiHsC changes their mass.
This force, F, is then, from Eq. 3:}

\textcolor{black}{
\begin{equation}
F=-c\frac{\partial m}{\partial t}
\end{equation}
}

\textcolor{black}{So that}

\textcolor{black}{
\begin{equation}
F=-c\frac{\partial m}{\partial x}\frac{\partial x}{\partial t}=-c^{2}\frac{\partial m}{\partial x}
\end{equation}
}

\textcolor{black}{Normally, of course, photons are not supposed to
have inertial mass in this way, but here this is assumed. It is not
clear what the size of this mass is, but it is clear for example that
light inside a mirrored box produces a kind of inertial mass for the
box. It is also assumed that the inertial mass of the microwave photons
(whatever its absolute value) is affected by MiHsC, but instead of
the horizon being the far-off and spherically symmetric Hubble horizon
as before, the horizon is now made by the asymmetric walls of the
cavity. This is possible because the photons involved are travelling
at the speed of light and are bouncing very fast between the two ends
of seperation 'L' and their acceleration ($a\sim v^{2}/L$) is so
large that the Unruh waves that are assumed in MiHsC to produce their
inertial mass are about the same size as the cavity, so they can be
affected by its walls, unlike the Unruh waves for a normal acceleration
which would be far to long to be affected by the cavity. This dependence
of the inertial mass on the width of the cavity means that the inertial
mass, is corrected by a MiHsC-like factor (Eq. 1) so the force in
Eq. 5 is modified as follows}

\textcolor{black}{
\begin{equation}
F=-c^{2}\left(\frac{m_{bigend}-m_{smallend}}{L}\right)
\end{equation}
}

\textcolor{black}{where L is the axial length of the cavity. In McCulloch
(2015) it was assumed that the waves only resonate perpendicular to
the axis. This model was quite successful in predicting the emdrive
thrusts (see the Table). MiHsC (Eq. 1) makes the assumption that the
Unruh wavelengths are made up of a complete pseudo-Planckian spectrum,
which is subsampled because some of the wavelengths do not fit within
the cavity and so the energy in the spectrum decays linearly as the
cavity narrows. Using Eq. 1, Eq. 6 becomes}

\textcolor{black}{
\begin{equation}
F=-\frac{mc^{2}}{L}\left(\frac{\lambda}{4w_{s}}-\frac{\lambda}{4w_{b}}\right)
\end{equation}
}

\textcolor{black}{This formula is not valid for pointed emdrive cones
where for example $w_{s}=0$, and to remedy this we need to consider
resonance along the axis. To do this it is necessary to calculate
the average distance between a photon at either end and the walls.
We can do this in a simplified manner by looking at this distance
along six orthogonal directions. Looking at Figure 1 and the distance
from the centre of the left-hand end plate (denoted P) to the walls:
these six directions are: to the left where the distance between P
and the wall is zero, to the right where it is L, up and down and
into and out of the page where it is $w_{s}$. This means that the
(very approximate) average distance the Unruh waves have to resonate
in at the narrow end are}

\textcolor{black}{
\begin{equation}
\bar{w}_{s}=\frac{0+L+4w_{s}}{6}
\end{equation}
}

\textcolor{black}{and at the wide end}

\textcolor{black}{
\begin{equation}
\bar{w_{b}=\frac{0+L+4w_{b}}{6}}
\end{equation}
}

\textcolor{black}{Substituting these into Eq. 7 we get}

\textcolor{black}{
\begin{equation}
F=-\frac{3mc^{2}\lambda}{2L}\left(\frac{1}{L+4w_{s}}-\frac{1}{L+4w_{b}}\right)
\end{equation}
}

\textcolor{black}{Here $\lambda$ is the wavelength of the Unruh radiation
seen by the microwaves because they are being reflected back and forth
by the cavity. This is given by$\lambda=8c^{2}/a=8c^{2}/(2c/(L/c))=4L$
so that}

\textcolor{black}{
\begin{equation}
F=-6mc^{2}\left(\frac{1}{L+4w_{s}}-\frac{1}{L+4w_{b}}\right)
\end{equation}
}

\textcolor{black}{Using $E=mc^{2}$ and $E=\int Pdt$ where P is the
power input into the cavity, gives}

\textcolor{black}{
\begin{equation}
F=-6\int Pdt\left(\frac{1}{L+4w_{s}}-\frac{1}{L+4w_{b}}\right)
\end{equation}
}

\textcolor{black}{Integrating P over one cycle (one trip of the microwaves
from end to end) gives Pt where t is the time taken for the microwaves
to travel from one end of the cavity's long axis to the other, which
is L/c, so}

\textcolor{black}{
\begin{equation}
F=-\frac{6PL}{c}\left(\frac{1}{L+4w_{s}}-\frac{1}{L+4w_{b}}\right)
\end{equation}
}

\textcolor{black}{This is for one trip along the cavity, but the Q
factor quantifies how many trips there are before the photon dissipates
so we need to multiply by Q. Note that the force is towards the narrow
end for both directions of travel as explained in Figure 1 and the
discussion.}

\textcolor{black}{
\begin{equation}
F=-\frac{6PQL}{c}\left(\frac{1}{L+4w_{s}}-\frac{1}{L+4w_{b}}\right)
\end{equation}
}

\textcolor{black}{Therefore MiHsC predicts that a new force will appear
acting always towards the narrow end of the cavity. This formula has
the advantage over that of McCulloch (2015) because it models, crudely,
the resonance of the Unruh waves in three-dimensions.}

\section{\textcolor{black}{Results}}

\textcolor{black}{The table includes a summary of the various experimental
results from Shawyer (2008) in rows 1 and 2 (denoted S1 and S2), the
Cannae drive in row 3 (Fetta, 2012) denoted C1, and Brady et al. (2014)
in rows 4 to 7 (denoted B1-B3 and B4v is the recent vacuum test) and
the vacuum test by Tajmar and Fiedler (2015) denoted T1. The Juan
et al. (2012) data is excluded because they did not specify their
Q factor or the geometry of their emdrive.}

\textcolor{black}{Column 1 names the experiment (S for Shawyer's experiment,
C for the Cannae drive, B for the NASA results and T for Tajmar and
Fiedler). Column 2 shows the input power (in Watts). Column 3 shows
the Q factor (dimensionless). Column 4 shows the axial length of the
cavity. Column 5 shows the width of the big and small ends (metres).
Column 6 shows the thrust predicted by one-dimensional MiHsC (Eq.
2) (in milliNewtons), column 7 shows the prediction of the three-dimensional
version of MiHsC derived in this paper (Eq. 14) and column 7 shows
the thrust observed (milli-Newtons) in the experiment.}

\textcolor{black}{}%
\begin{tabular}{|c|c|c|c|c|c|c|c|}
\hline 
\textcolor{black}{Expt} & \textcolor{black}{P } & \textcolor{black}{Q} & \textcolor{black}{L} & \textcolor{black}{$w_{big}/w_{small}$ } & \textcolor{black}{$F_{1d}$} & \textcolor{black}{$F_{3d}$} & \textcolor{black}{$F_{Obs}$}\tabularnewline
\hline 
\hline 
 & \textcolor{black}{W} &  & \textcolor{black}{m} & \textcolor{black}{metres} & \textcolor{black}{$mN$} & \textcolor{black}{$mN$} & \textcolor{black}{$mN$}\tabularnewline
\hline 
\textcolor{black}{S1} & \textcolor{black}{850} & \textcolor{black}{5900} & \textcolor{black}{0.156} & \textcolor{black}{0.16/0.1275} & \textcolor{black}{4.2} & \textcolor{black}{3.8} & \textcolor{black}{16}\tabularnewline
\hline 
\textcolor{black}{S2} & \textcolor{black}{1000} & \textcolor{black}{45,000} & \textcolor{black}{0.345} & \textcolor{black}{0.28/0.1289} & \textcolor{black}{217} & \textcolor{black}{149} & \textcolor{black}{80-214}\tabularnewline
\hline 
\textcolor{black}{C1} & \textcolor{black}{10.5} & \textcolor{black}{$11\times10^{6}$} & \textcolor{black}{0.03} & \textcolor{black}{0.22/0.2} & \textcolor{black}{5.3} & \textcolor{black}{7.3} & \textcolor{black}{9}\tabularnewline
\hline 
\textcolor{black}{B1} & \textcolor{black}{16.9} & \textcolor{black}{7,320} & \textcolor{black}{0.2286} & \textcolor{black}{0.2794/0.1588} & \textcolor{black}{0.26} & \textcolor{black}{0.23} & \textcolor{black}{0.091}\tabularnewline
\hline 
\textcolor{black}{B2} & \textcolor{black}{16.7} & \textcolor{black}{18,100} & \textcolor{black}{``} & \textcolor{black}{``} & \textcolor{black}{0.63} & \textcolor{black}{0.57} & \textcolor{black}{0.05}\tabularnewline
\hline 
\textcolor{black}{B3} & \textcolor{black}{2.6} & \textcolor{black}{22,000} & \textcolor{black}{``} & \textcolor{black}{``} & \textcolor{black}{0.12} & \textcolor{black}{0.11} & \textcolor{black}{0.055}\tabularnewline
\hline 
\textcolor{black}{B4v} & \textcolor{black}{50} & \textcolor{black}{6730} & \textcolor{black}{``} & \textcolor{black}{``} & \textcolor{black}{0.70} & \textcolor{black}{0.64} & \textcolor{black}{0.03}\tabularnewline
\hline 
\textcolor{black}{T1} & \textcolor{black}{700} & \textcolor{black}{20} & \textcolor{black}{0.1008} & \textcolor{black}{0.1062/0.075} & \textcolor{black}{0.02} & \textcolor{black}{0.02} & \textcolor{black}{0.02-0.11}\tabularnewline
\hline 
\end{tabular}

\textcolor{black}{Table 1. A summary of the fully documented emdrive
experiments so far. Column 1 shows the experiment name, column 2 shows
the input power, column 3 the Q factor, column 4 the cavity's axial
length, column 5 shows the cavity end widths. Columns 6 and 7 show
the thrusts predicted by the 1-d and 3-d versions of MiHsC respectively,
and column 8 shows the observed thrust.}

\textcolor{black}{The Table shows that MiHsC predicts the experimental
results quite well. The worst agreements are for B2 and B4v where
MiHsC is over a factor of ten out. In almost all cases MiHsC3d performs
slightly better than MiHsC1d. The differences could be due to the
approximate way MiHsC has been applied so far and it is unclear what
the error bars on the observations are. The range of values for S2
gives some idea of the large uncertainty in the data. However, MiHsC
predicts the correct order of magnitude for all the cases. This is
encouraging given that this model is rather approximate (not fully
three-dimensional), may be affected by uncertainties in the cavity
geometry. It should also be noted that MiHsC has no adjustable parameters.}

\section{\textcolor{black}{Discussion}}

\textcolor{black}{To explain in a more intuitive manner: MiHsC predicts
that as photons travel from the narrow end on the left of the emdrive
to the wide end on the right (see the lower arrows in Figure 1) their
inertial mass increases as more Unruh wavelengths fit at the righthand
wide end of the cavity. In the figure this is shown by the arrow being
thicker on the right. This change though, has violated the conservation
of momentum, so we must now apply a force towards the narrow end to
slow the photon down and conserve momentum. When the photons bounce
off the wide end and move leftwards again towards the narrow end they
lose inertial mass because fewer Unruh waves fit at the narrow end,
so to conserve momentum it is now necessary to apply a force, again
towards the narrow end, to speed the photons up. In both cases the
new MiHsCian force is towards the narrow end and of a size, as shown
in the Table, similar to the anomalous thrust that has been seen in
the experiments.}

\textcolor{black}{More data is needed for comparison, and a more accurate
modelling of the effects of MiHsC will be needed. This analysis for
simplicity, assumed the microwaves only travelled along the axis and
the three-dimensional resonance of the waves was only crudely modelled:
a full 3-d model is needed.}

\textcolor{black}{This proposal predicts the observations quite well,
but makes two controversial assumptions. For example that the inertial
mass of photons is finite (in defence of this, they do carry momentum)
and varies in line with MiHsC, and that the speed of the light is
changing in the cavity. So it is important to suggest a definite test.}

\textcolor{black}{Both Eq. 14 and the simpler equation in McCulloch
(2015), Eq. 2, predict that it should be possible to reverse the sign
of the thrust by shortening the usual cavity length (L) or changing
the frequency so that the Unruh waves fit better into the short end
($w_{s}$) than the wide end ($w_{b}$). This thrust reversal may
have been seen in recent NASA experiments. Equation 14 also suggests
that the anomalous force can be increased by increasing the power
input, P, or the quality factor of the cavity (Q, the number of times
the microwaves bounce between the two ends) and the speed of light
on the denominator of Eq. 14 implies that if the value of c was decreased
by a dielectric the effect would be enhanced.}

\section{\textcolor{black}{Conclusion}}

\textcolor{black}{More than eight tests in four independent labs have
shown that when microwaves resonate within an asymmetric cavity an
anomalous thrust is generated pushing the cavity towards its narrow
end.}

\textcolor{black}{This force can be predicted fairly well by using
a new model for inertia (MiHsC) which assumes that the inertial mass
of the photons is caused by Unruh radiation whose wavelengths have
to fit exactly inside the cavity so that the photons' inertial mass
is greater at the wide end. To conserve momentum a new force appears
to to push the cavity towards its narrow end, and the predicted force
is similar to the thrust observed.}

\textcolor{black}{MiHsC suggests that the thrust can be increased
by increasing the input power, the Q factor, or using a dielectric.
As a direct test MiHsC predicts that the thrust can be reversed by
making the length L equal to the width of the narrow end.}

\section*{\textcolor{black}{Acknowledgements}}

\textcolor{black}{Many thanks for Dr Jose Rodal and others on an NSF
forum for estimating from photographs the proportions of the various
experimental arrangements.}

\section*{\textcolor{black}{References}}

\textcolor{black}{Brady, D.A., H.G. White, P. March, J.T. Lawrence
and F.J. Davies, 2014. Anomalous thrust production from an RF test
device measured on a low-thrust torsion pendulum. 50th AIAA/ASME/SAE/ASEE
Joint Propulsion conference.}

\textcolor{black}{Fetta, G.P., 2014. Numerical and experimental results
for a novel propulsion technology requiring no on-board propellent.
50th AIAA/ASME/SAE/ASEE Joint Propulsion Conference. AIAA.}

\textcolor{black}{Juan, Y., 2012. Net thrust measurement of propellantless
microwave thrusters. Acta Physica Sinica, 61, 11.}

\textcolor{black}{McCulloch, M.E., 2007. The Pioneer anomaly as modified
inertia. MNRAS, 376, 338.}

\textcolor{black}{McCulloch, M.E., 2010. Minimum accelerations from
quantised inertia. EPL, 90, 29001.}

\textcolor{black}{McCulloch, M.E., 2012. Testing quantised inertia
on galactic scales. Astro. \& Space Sci., 342, 575.}

\textcolor{black}{McCulloch, M.E., 2015. Can the emdrive be explained
by quantised inertia? Progress in Physics, 11, 78-80.}

\textcolor{black}{Mullins, J., 2006. Relativity drive: the end of
wings and wheels? New Scientist (2568), p30-34.}

\textcolor{black}{Shawyer, R, 2008. Microwave propulsion - progress
in the emdrive programme. 59th International Astronautical conference.
IAC-2008. Glasgow, UK.}

\textcolor{black}{Tajmar, M., G. Fiedler, 2015. Direct thrust measurements
of an EM drive and evaluation of possible side effects. 51st AIAA/SAE/ASEE
Joint Propulsion Conference, Propulsion and Energy Forum, Orlando,
Florida.}

\section*{\textcolor{black}{Figures}}

\textcolor{black}{\includegraphics{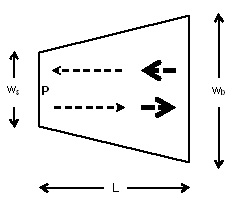}}

\textcolor{black}{Figure 1. The emdrive cavity (solid line) showing
a photon moving right (lower dashed arrows) and then bouncing back
towards the left (upper arrows). The photon's mass is shown by line
thickness and its speed is shown by the line length. For both rightward
and leftward moving photons the mass change caused by MiHsC violates
the conservation of momentum, which can only be satisfied by, in both
cases, a force acting towards the narrow end, changing its speed.}
\end{document}